\documentclass[conference]{IEEEtran}
\usepackage[algoruled, linesnumbered]{algorithm2e}
\usepackage{cite}
\usepackage{graphicx}
\usepackage{psfrag}
\usepackage{epsfig}
\usepackage{subfigure}
\usepackage{url}
\usepackage{stfloats}
\usepackage{amsmath}
\usepackage{array}
\usepackage{array}
\begin{document}
\title{Improving Performance of Speaker Identification System Using Complementary Information Fusion}
\author{\authorblockN{Md. Sahidullah, Sandipan Chakroborty and
Goutam Saha\\
\authorblockA{Department of Electronics and Electrical Communication
Engineering\\Indian Institute of Technology, Kharagpur, India,
Kharagpur-721 302\\Email: sahidullah@iitkgp.ac.in, mail2sandi@gmail.com, gsaha@ece.iitkgp.ernet.in
\\Telephone:
+91-3222-283556/1470, FAX: +91-3222-255303}}
\\
}

\maketitle

\begin{abstract}
Feature extraction plays an important role as a front-end processing block in speaker identification (SI) process. Most of the SI systems utilize like Mel-Frequency Cepstral Coefficients (MFCC), Perceptual Linear Prediction (PLP), Linear Predictive Cepstral Coefficients (LPCC), as a feature for representing speech signal. Their derivations are based on short term processing of speech signal and they try to capture the vocal tract information ignoring the contribution from the vocal cord. Vocal cord cues are equally important in SI context, as the information like pitch frequency, phase in the residual signal, etc could convey important speaker specific attributes and are complementary to the information contained in spectral feature sets. In this paper we propose a novel feature set extracted from the residual signal of LP modeling. Higher-order statistical moments are used here to find the nonlinear relationship in residual signal. To get the advantages of complementarity vocal cord based decision score is fused with the vocal tract based score. The experimental results on two public databases show that fused mode system outperforms single spectral features.
\end{abstract}

\section{Introduction}
Speaker Identification is the process of identifying a person by his/her voice signal \cite{camp}. A state-of-the art speaker identification system requires feature extraction unit as a front end processing block followed by an efficient modeling scheme. Vocal tract information like its formant frequency, bandwidth of formant frequency etc. are supposed to be unique for human beings. The basic target of the feature extraction block is to characterize those information. On the other hand this feature extraction process represents the original speech signal into a compact format as well as emphasizing the speaker specific information. The function of the feature extraction process block is also to represent the original signal into a robust manner. Most of the speaker identification system uses Mel Frequency Cepstral coefficients (MFCC) or Linear Prediction Cepstral Coefficient (LPCC) as a feature extraction block \cite{camp}. MFCC is the modification of conventional Linear Frequency Cepstral Coefficient keeping in mind the auditory system of human being \cite{davis}. On the other hand, the LPCC is based on time domain processing of speech signal \cite{atal}. Later conventional LPCC is also modified motivated by perceptual property of human ear \cite{hermansky}.
Like vocal tract, Vocal cord information also contains some speaker specific information \cite{mahadeva}. Residual signal which can be obtained from the Linear Prediction (LP) analysis of speech signal contains information related to source or vocal cord. Earlier Auto-associative Neural Network (AANN), Wavelet Octave Coefficients of Residues (WOCOR), residual phase etc. were used to extract the information from residual signal. In this work we have introduced Higher-order Statistical Moments to capture the information from the residual signal. In this paper we are integrating the vocal cord information with vocal tract information to boost up the performance of speaker identification system. The log likelihood score of both the system are fused together to get the advantages of their complementarity \cite{ksrmurty,nengheng}. The speaker identification results on both the databases prove that combining the two systems, the performance can be improved over baseline spectral feature based systems.
\par
This paper is organized as follows. In section II we first review the basic of linear prediction analysis followed by the proposed feature extraction technique. The speaker identification experiment with results is shown in section III. Finally, the paper is concluded in section IV.

\section{Feature Extraction From Residual Signal}
In this section we first explain the conventional method of derivation of residual signal by LP-analysis. The proposed feature extraction process is described consequently.

\begin{figure*}
\centering
\includegraphics[height =12 cm, width=18 cm]{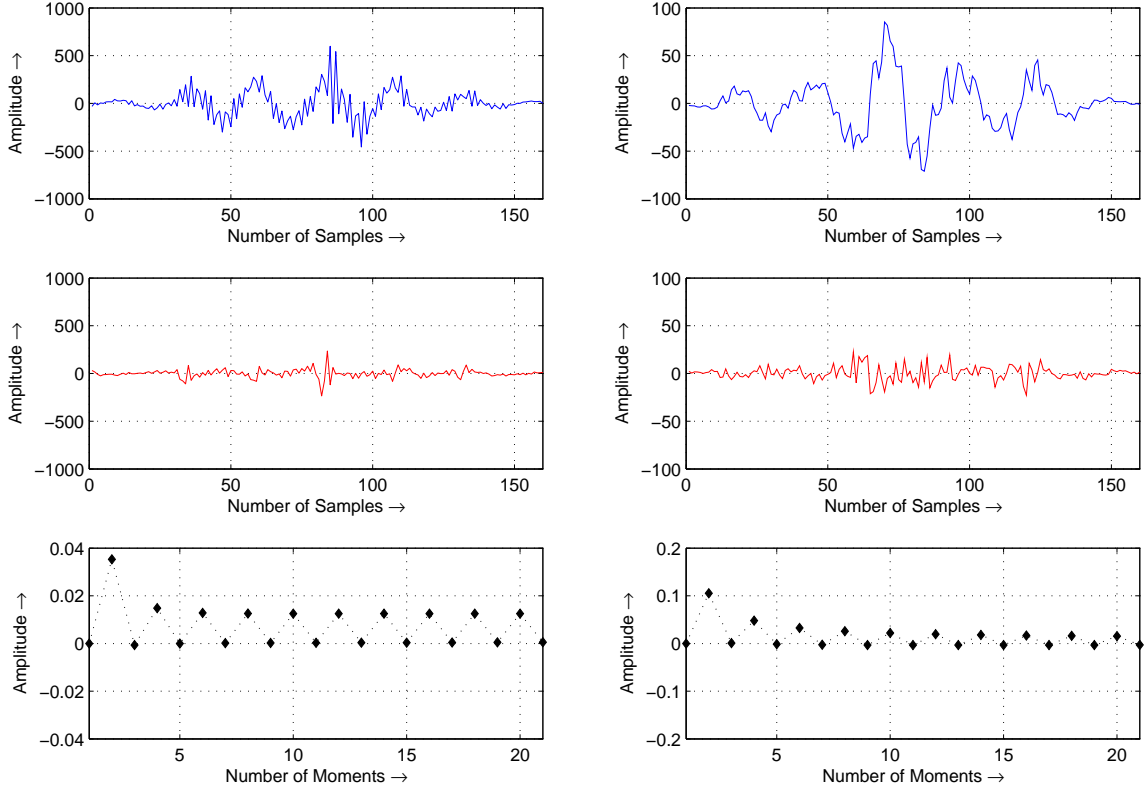}\\
\caption{Example of two speech frames (top), their LP residuals (middle) and corresponding residual moments (bottom).}
\label{plotmoment}
\centering
\end{figure*}

\subsection{Linear Prediction Analysis and Residual Signal}\label{lpccomp}
In the LP model, $(n-1)$-th to $(n-p)$-th samples of the speech wave ($n$, $p$ are integers) are used to predict the $n$-th sample. The predicted value of the $n$-th speech sample \cite{atal} is given by

\begin{equation}\label{lp1}
\hat{s}(n)=\sum_{k=1}^{p}a(k)s(n-k)
\end{equation}

where $\{a(k)\}_{k=1}^{p}$ are the predictor coefficients and $s(n)$ is the $n$-th speech sample.The value of $p$ is chosen such that it could effectively capture the real and complex poles of the vocal tract in a frequency range equal to half the sampling frequency.The Prediction Coefficients (PC) are determined by minimizing the mean square prediction error \cite{camp} and the error is defined as
\begin{equation}\label{lp2}
E(n)=   \frac{1}{N}\sum_{n=0}^{N-1}(s(n)-\hat{s}(n))^{2}
\end{equation}
where summation is taken over all samples i.e., $N$.
The set of coefficients $\{a(k)\}_{k=1}^{p}$ which minimize the mean-squared prediction error are obtained as solutions of the set of linear equation
\begin{equation}\label{lp3}
\sum_{k=1}^{p}\phi(j,k)a(k)=\phi(j,0) , j=1,2,3,\ldots,p
\end{equation}
where
\begin{equation}\label{lp4}
\phi(j,k)=\frac{1}{N}\sum_{n=0}^{N-1}s(n-j)s(n-k)
\end{equation}

The PC, $\{a(k)\}_{k=1}^{p}$ are derived by solving the recursive equation (\ref{lp3}).

Using the $\{a(k)\}_{k=1}^{p}$ as model parameters, equation (\ref{lp5}) represents the fundamental basis of LP representation. It implies that any signal can be defined by a linear predictor and its prediction error.

\begin{equation}\label{lp5}
s(n)=-\sum_{k=1}^{p}a(k)s(n-k)+e(n)
\end{equation}

The LP transfer function can be defined as,
\begin{equation}\label{lp6}
H(z)=\frac{G}{1 + \sum_{k=1}^{p}a(k)z^{-k}}=\frac{G}{A(z)}
\end{equation}
where $G$ is the gain scaling factor for the present input and $A(z)$ is the $p$-th order inverse filter.
These LP coefficients itself can be used for speaker recognition as it contains some speaker specific information like vocal tract resonance frequencies, their bandwidths etc.

The prediction error i.e., $e(n)$ is called Residual Signal and it contains all the complementary information that are not contained in the PC. Its worth mentioning here that residual signal conveys vocal source cues containing fundamental frequency, pitch period etc.

\subsection{Statistical Moments of Residual Signal}\label{hosmr}
Residual signal which is introduced in Section \ref{lpccomp} generally has a noise like behavior and it has flat spectral response. Though it contains vocal source information, it is very difficult to perfectly characterize it. In literature Wavelet Octave Coefficients of Residues (WOCOR) \cite{nengheng}, Auto-associative Neural Network (AANN) \cite{mahadeva} , residual phase \cite{ksrmurty} etc are used to extract the residual information. It is worth mentioning here that higher-order statistics have shown significant results in a number of signal processing applications \cite{nandi} when the nature of the signal is non-gaussian.  Higher order statistics also got attention of the researchers for retrieving information from the LP residual signals \cite{nemer}. Recently, higher order cumulant of LP residual signal is investigated \cite{chetouani} for improving the performance of speaker identification system.
\par
Higher order statistical moments of a signal parameterizes the shape of a function \cite{lo}.
Let the distribution of random signal $x$ be denoted by $P(x)$, the central moment of order $k$ of $x$ be denoted by
\begin{equation}\label{eqmoment}
M_k  = \int\limits_{ - \infty }^\infty  {(x - \mu )^k } dP
\end{equation}
for $k=1,2,3...$, where $\mu$ is the mean of $x$.

On the other hand, the characteristics function of the probability distribution of the random variable is given by,

\begin{equation}\label{eqchs}
\varphi _X (t) = \int\limits_{ - \infty }^\infty  {e^{jtx} dP = \sum\limits_{k = 0}^\infty  {M_k \frac{{(jt)^k }}{{k!}}} }
\end{equation}
From the above equation it is clear that moments ($M_k$) are coefficients for the expansion of the characteristics function. Hence, they can be treated as one set of expressive constants of a distribution. Moments can also effectively capture the randomness of residual signal of auto regressive modeling \cite{mattson}.
\par
\begin{figure}
\centering
\includegraphics[height =4 cm, width=6 cm]{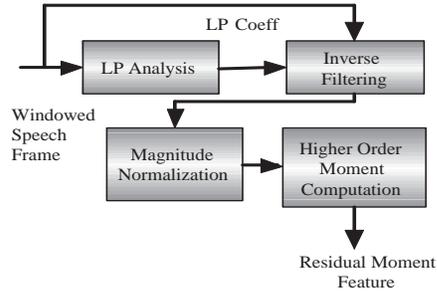}\\
\caption{Block diagram of Residual Moment Based Feature Extraction Technique.}\label{bdmoment}
\centering
\end{figure}

In this paper, we use higher order statistical moments of residual signal to parameterize the vocal source information. The feature derived by the proposed technique is termed as Higher Order Statistical Moment of Residual (HOSMR). The different blocks of the proposed feature extraction technique from residual are shown in fig. \ref{bdmoment}.

At first the residual signal is first normalized between the range $[-1,+1]$.
Then central moment of order $k$ of a residual signal $e(n)$ is computed as,
\begin{equation}\label{moment}
m_{k}=\frac{1}{N}\sum_{n=0}^{N-1}(e(n)-\mu)^{k}
\end{equation}
where, $\mu$ is the mean of residual signal over a frame. As the range of the residual signal is normalized, the first order moment (i.e. the mean) becomes zero. The higher order moments (for $k=2,3,4...K$) are taken as vocal source features as they represent the shape of the distribution of random signal. The lower order moments are coarse parametrization whereas the higher orders are finer representation of residual signal. In fig. \ref{plotmoment}, LP residual signal of a frame is shown as well as its higher order moments. It is clear from the picture that if the lower order moments are considered both the even and odd order values are highly differentiable.

\subsection{Fusion of Vocal Tract and Vocal Cord Information}\label{fusion}
\begin{figure*}
\centering
\includegraphics[height =12 cm, width=14 cm]{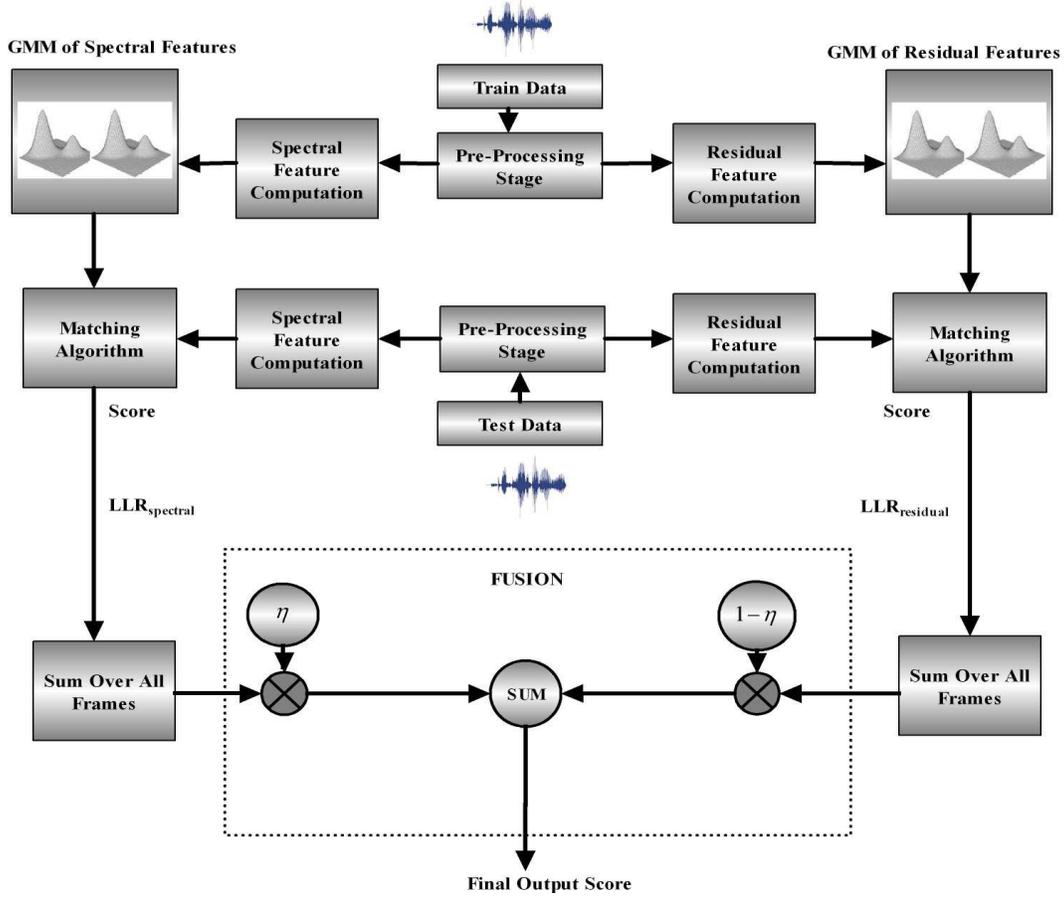}\\
\caption{Block diagram of Fusion Technique: Score level fusion of Vocal tract (short term spectral based feature) and Vocal cord information (Residual).}\label{bdfusion}
\centering
\end{figure*}

In this section we propose to integrate vocal tract and vocal cord parameters identifying speakers. In spite of the two approaches have significant performance difference, the way they represent speech signal is complementary to one another. Hence, it is expected that combining the advantages of both the feature will improve \cite{kittler} the overall performance of speaker identification system. The block diagram of the combined system is shown in fig. \ref{bdfusion}. Spectral features and Residual features are extracted from the training data in two separate streams.
Consequently, speaker modeling is performed for the respective features independently and model parameters are stored in the model database. At the time of testing same process is adopted
for feature extraction. Log-likelihood of two different features are computed w.r.t. their corresponding models. Finally, the output score is weighted and combined.

We have used score level linear fusion which can be formulated as in equation (\ref{fusioneq}). To get the advantages of both the system and their complementarity the score level linear fusion can be formulated as follows:
\begin{equation}\label{fusioneq}
    LLR_{combined}=\eta LLR_{spectral}+(1-\eta) LLR_{residual}
\end{equation}
where $LLR_{spectral}$ and $LLR_{residual}$ are log-likelihood ratio calculated from the spectral and residual based systems, respectively. The fusion weight is decided by the parameter $\eta$.

\section{Speaker Identification Experiment}
\subsection{Experimental Setup}\label{ee}

\subsubsection{Pre-processing stage}
In this work, pre-processing stage is kept similar throughout different features extraction methods. It is performed using the following steps:
\begin{itemize}
  \item Silence removal and end-point detection are done using energy threshold criterion.
  \item The speech signal is then pre-emphasized with $0.97$ pre-emphasis factor.
  \item The pre-emphasized speech signal is segmented into frames of each $20\mathrm{ms}$ with $50\%$ overlapping ,i.e. total number of samples in each frame is $N = 160$, (sampling frequency $F_s = 8KHz$.
  \item In the last step of pre-processing, each frame is windowed using hamming window given equation
\begin{equation}\label{hamming}
w(n)=0.54-0.46\cos(\frac{2\pi n}{N-1})
\end{equation}
where $N$ is the length of the window.
\end{itemize}

\subsubsection{Classification \& Identification stage}
Gaussian Mixture Modeling (GMM) technique is used to get probabilistic model for the feature vectors of a speaker. The idea of GMM is to use weighted summation of multivariate gaussian functions to represent the probability density of feature vectors and it is given by

\begin{equation}\label{gmm1}
p(\textbf{x})=\sum_{i=1}^{M}p_i b_i(\textbf{x})
\end{equation}
where $\textbf{x}$ is a $d$-dimensional feature vector, $b_i(\textbf{x})$, $i=1,...,M$ are the  component densities and $p_i$, $i = 1,...,M$ are the mixture weights or $\emph{prior}$ of individual gaussian. Each component density  is given by
\begin{equation}\label{gmm2}
b_i(\textbf{x})=\frac{1}{(2\pi)^{\frac{d}{2}}|\mathbf{\Sigma_{i}}|^{\frac{1}{2}}}\exp \bigg\{ -\frac{1}{2}( \textbf{x} - \boldsymbol{\mu_i)}^{t}\mathbf{\Sigma_{i}}^{-1}( \textbf{x} - \boldsymbol{\mu_i})\bigg\}
\end{equation}
with mean vector $\boldsymbol{\mu_i}$ and covariance matrix $\mathbf{\Sigma_{i}}$. The mixture weights must satisfy the constraint that $\sum_{i=1}^{M}p_{i}=1$ and $p_{i}$ $\geq$ $0$.
The Gaussian Mixture Model is parameterized by the mean, covariance and mixture weights from all component densities and is denoted by
\begin{equation}\label{gmm3}
\lambda=\{ p_{i},\boldsymbol{\mu_i},\mathbf{\Sigma_{i}} \}_{i=1}^{M}
\end{equation}
In SI, each speaker is represented by the a GMM and is referred to by his/her model $\lambda$. The parameter of $\lambda$ are optimized using Expectation Maximization(EM) algorithm \cite{demp}. In these experiments, the GMMs are trained with 10 iterations where clusters are initialized by vector quantization \cite{lbg} algorithm.

In identification stage, the log-likelihood scores of the feature vector of the utterance under test is calculated by
\begin{equation}\label{gmm4}
\log p(\boldsymbol{X}|\lambda)=\sum_{t=1}^{T}p(\boldsymbol{x}_t|\lambda)
\end{equation}
Where $\textbf{X}=\{\boldsymbol{x}_1,\boldsymbol{x}_2,...,\boldsymbol{x}_t \}$ is the feature vector of the test utterance.

In closed set SI task, an unknown utterance is identified as an utterance of a particular speaker whose model gives maximum log-likelihood.
It can be written as
\begin{equation}\label{gmm5}
\hat{S}=\arg \mathop {\max }\limits_{1 \le k \le S}\sum_{t=1}^{T}p(\boldsymbol{x}_t|\lambda_{k})
\end{equation}
where $\hat{S}$ is the identified speaker from speaker's model set $\Lambda =\{\lambda_1,\lambda_2,...,\lambda_S\}$ and $S$ is the total number of speakers.

\subsubsection{Databases for experiments}

\hspace{1cm}

\emph{YOHO Database:} The YOHO voice verification corpus \cite{camp,yoho} was collected while
testing ITT's prototype speaker verification system in an office
environment. Most subjects were from the New York City area,
although there were many exceptions, including some non-native
English speakers. A high-quality telephone handset (Shure XTH-383)
was used to collect the speech; however, the speech was not passed
through a telephone channel. There are $138$ speakers ($106$ males
and $32$ females); for each speaker, there are $4$ enrollment
sessions of $24$ utterances each and $10$ test sessions of $4$
utterances each. In this work, a closed set text-independent
speaker identification problem is attempted where we consider all
$138$ speakers as client speakers. For a speaker, all the $96\ (4\
\mathrm{sessions}\times24\ \mathrm{utterances})$ utterances are
used for developing the speaker model while for testing, $40\ (10\
\mathrm{sessions}\times 4 \ \mathrm{utterances})$ utterances are
put under test. Therefore, for $138$ speakers we put
$138\times40=5520$ utterances under test and evaluated the
identification accuracies.
\par
\emph{POLYCOST Database:} The POLYCOST database \cite{melin} was recorded as a common
initiative within the COST $250$ action during January- March
1996. It contains around $10$ sessions recorded by $134$ subjects
from $14$ countries. Each session consists of $14$ items, two of
which (MOT01 \& MOT02 files) contain speech in the subject's
mother tongue. The database was collected through the European
telephone network. The recording has been performed with ISDN
cards on two XTL SUN platforms with an $8$ kHz sampling rate. In
this work, a closed set text independent speaker identification
problem is addressed where only the mother tongue (MOT) files are
used. Specified guideline \cite{melin} for conducting closed set
speaker identification experiments is adhered to, i.e. `MOT02'
files from first four sessions are used to build a speaker model
while `MOT01' files from session five onwards are taken for
testing. As with YOHO database, all speakers ($131$ after deletion
of three speakers) in the database were registered as clients.

\subsubsection{Score Calculation}
In closed-set speaker identification problem, identification
accuracy as defined in \cite{rey} and given by the equation (\ref{eqpia}) is followed.
\begin{eqnarray}\label{eqpia}
\mathrm{Percentage\ of\ identification\ accuracy\
(\textbf{\textbf{PIA}})}&=&\nonumber\\
\frac{\mathrm{No.\ of\ utterance\ correctly\
identified}}{\mathrm{Total\ no.\ of\ utterance\ under\
test}}\times 100\label{poc}
\end{eqnarray}

\subsection{Speaker Identification Experiments and Results}
The performance of speaker identification system based on the proposed HOSMR feature is evaluated on both the databases. The order of LP is kept at $17$ and $6$ residual moments are taken to characterize the residual information. We have conducted experiment based on GMM based classifier for different model order. The identification results are shown in Table \ref{tablehosmr}. The identification performance is very low because the vocal cord parameters are not the only cues for identifying speakers but it has some inherent contribution in recognition. At the same time it contains information which are not contained in spectral feature. The combined performance of both the system is to be observed. We have conducted SI experiment using two major kinds of baseline features, some are based on LP analysis (LPCC and PLPCC) and others (LFCC and MFCC) are based on filterbank analysis. The feature dimension is set at $19$ for all kinds of features for better comparison. In LP based systems $19$ filters are used for all-pole modeling of speech signals. On the other hand $20$ filters are used for filterbank based system and $19$ coefficients are taken for extracting Linear Frequency Cepstral Coefficients (LFCC) and MFCC after discarding the first co-efficient which represents dc component. The detail description are available in \cite{sandi1,sandi}. The derivation LP based features can be found in \cite{camp,rab,hermansky}.

\par

The performance of baseline SI systems and fused systems for different features and different model orders are shown in Table \ref{tablepoly} and Table \ref{tableyoho} for POLYCOST and YOHO databases respectively.  In this experiment, we take equal evidence from the two systems and set the value of $\eta$ to $0.5$. The results for the conventional spectral features follows the results shown in \cite{rey1}. The POLYCOST database consists of speech signals collected over telephone channel. The improvement for this database is significant over the YOHO which is micro-phonic. The experimental results shows significant performance improvement for SI system compare to only spectral systems for various model order.

\begin{table}[htbp]
\begin{center}
\caption{Speaker Identification Results on POLYCOST and YOHO database using HOSMR feature for different model order of GMM (HOSMR Configuration: LP Order $=17$, Number of Higher Order Moments$=6$).}
\begin{tabular}{|c|c|c|}
\hline
\hline
Database &
Model Order&
Identification Accuracy \\
\hline
\hline
\raisebox{-4.50ex}[0cm][0cm]{POLYCOST}&
2&  19.4960\\
\cline{2-3}
 & 4& 21.6180 \\
\cline{2-3}
 & 8& 19.0981 \\
\cline{2-3}
 & 16&  22.4138 \\
\hline
\hline
\raisebox{-4.50ex}[0cm][0cm]{YOHO}&
2&  16.8841\\
\cline{2-3}
 & 4&  18.2246\\
\cline{2-3}
 & 8&  15.1268\\
\cline{2-3}
 & 16&  18.2246\\
\cline{2-3}
& 32&  21.2138\\
\cline{2-3}
 & 64&  21.9565\\
\hline
\hline
\end{tabular}
\label{tablehosmr}
\end{center}
\end{table}

\begin{table}[htbp]
\begin{center}
\caption{Speaker Identification Results on POLYCOST database showing the performance of baseline (single stream) system and fused system (HOSMR Configuration: LP Order $=17$, Number of Higher Order Moments$=6$, Fusion Weight ($\eta$)$=0.5$).}
\begin{tabular}{|c|c|c|c|}
\hline
\hline
Feature &
Model Order&
Baseline System&
Fused System  \\
\hline
\hline
\raisebox{-4.50ex}[0cm][0cm]{LPCC}&
2& 63.5279& \textbf{71.4854} \\
\cline{2-4}
 & 4& 74.5358& \textbf{78.9125} \\
\cline{2-4}
 & 8& 80.3714& \textbf{81.6976} \\
\cline{2-4}
 & 16& 79.8408& \textbf{82.8912} \\
\hline
\hline
\raisebox{-4.50ex}[0cm][0cm]{PLPCC}&
2& 62.9973& \textbf{65.7825} \\
\cline{2-4}
 & 4& 72.2812& \textbf{75.5968} \\
\cline{2-4}
 & 8& 75.0663& \textbf{77.3210} \\
\cline{2-4}
 & 16& 78.3820& \textbf{80.5040} \\
\hline
\hline
\raisebox{-4.50ex}[0cm][0cm]{LFCC}&
2& 62.7321& \textbf{71.6180} \\
\cline{2-4}
 & 4& 74.9337& \textbf{78.1167} \\
\cline{2-4}
 & 8& 79.0451& \textbf{81.2997} \\
\cline{2-4}
 & 16& 80.7692& \textbf{83.4218} \\
\hline
\hline
\raisebox{-4.50ex}[0cm][0cm]{MFCC}&
2& 63.9257&\textbf{69.7613} \\
\cline{2-4}
 & 4& 72.9443& \textbf{76.1273} \\
\cline{2-4}
 & 8& 77.8515& \textbf{79.4430} \\
\cline{2-4}
 & 16& 77.8515& \textbf{79.5756} \\
\hline
\hline
\end{tabular}
\label{tablepoly}
\end{center}
\end{table}

\begin{table}[htbp]
\begin{center}
\caption{Speaker Identification Results on YOHO database showing the performance of baseline (single stream) system and fused system (HOSMR Configuration: LP Order $=17$, Number of Higher Order Moments$=6$, Fusion Weight ($\eta$)$=0.5$)).}
\begin{tabular}{|c|c|c|c|}
\hline
\hline
Feature& Model Order& Baseline System & Fused System  \\
\hline
\hline
\raisebox{-7.50ex}[0cm][0cm]{LPCC}&
2& 80.9420& \textbf{84.7101} \\
\cline{2-4}
& 4& 88.9855& \textbf{91.0870} \\
\cline{2-4}
 & 8& 93.8949& \textbf{94.7826} \\
\cline{2-4}
 & 16& 95.6884& \textbf{96.2862} \\
\cline{2-4}
 & 32& 96.5399& \textbf{97.1014} \\
\cline{2-4}
 & 64& 96.7391& \textbf{97.2826} \\
\hline
\hline
\raisebox{-7.50ex}[0cm][0cm]{PLPCC}&
2& 66.5761& \textbf{72.5543} \\
\cline{2-4}
 & 4& 76.9203& \textbf{81.0507} \\
\cline{2-4}
 & 8& 85.3080& \textbf{87.7717} \\
\cline{2-4}
 & 16& 90.6341& \textbf{91.9022} \\
\cline{2-4}
 & 32& 93.5326& \textbf{94.3116} \\
\cline{2-4}
 & 64& 94.6920& \textbf{95.3986} \\
\hline
\hline
\raisebox{-7.50ex}[0cm][0cm]{LFCC}&
2& 83.0072& \textbf{85.8152} \\
\cline{2-4}
 & 4& 90.3623&  \textbf{91.7935} \\
\cline{2-4}
 & 8& 94.6196& \textbf{95.4891} \\
\cline{2-4}
 & 16& 96.2681& \textbf{96.6848} \\
\cline{2-4}
 & 32& 97.1014& \textbf{97.3551} \\
\cline{2-4}
 & 64& 97.2464& \textbf{97.6268} \\
\hline
\hline
\raisebox{-7.50ex}[0cm][0cm]{MFCC}&
2& 74.3116& \textbf{78.6051} \\
\cline{2-4}
 & 4& 84.8551& \textbf{86.9384} \\
\cline{2-4}
 & 8& 90.6703& \textbf{92.0290} \\
\cline{2-4}
 & 16& 94.1667& \textbf{94.6920} \\
\cline{2-4}
 & 32& 95.6522& \textbf{95.9964} \\
\cline{2-4}
 & 64& 96.7935& \textbf{97.1014} \\
\hline
\hline
\end{tabular}
\label{tableyoho}
\end{center}
\end{table}

\section{Conclusion}

The objective of this paper is to propose a new technique to improve the performance of conventional speaker identification system which are based on spectral features representing only vocal tract information. Higher-order statistical moment of residual signal is derived and treated as a parameter carrying vocal cord information. The log likelihood of both the system are fused together. The experimental results on two popular speech corpus prove that significant improvement can be obtained in combined SI system.

\bibliography{latexbib}

\end{document}